\documentclass[aps,showpacs,pra,twocolumn,amssymb,amsmath]{revtex4}
\usepackage{graphicx}
\usepackage{dcolumn}
\usepackage{bm}

\begin{document}
\title{Exact Solution of a Yang-Baxter Spin-$1/2$ Chain Model and Quantum
Entanglement}
\author{Ming-Guang Hu$^1$}
 \email{huphys@hotmail.com}
\author{Kang Xue$^2$}
\author{Mo-Lin Ge$^1$}
\email{geml@nankai.edu.cn}
\affiliation{$^1$Theoretical Physics
Division, Chern Institute of Mathematics, Nankai University, Tianjin
300071, P. R.
China\\
$^2$Department of Physics, Northeast Normal University, Changchun,
Jilin 130024, P. R. China}

\date{\today}

\begin{abstract}
Entanglement is believed to be crucial in macroscopic physical
systems for understanding the collective quantum phenomena such as
quantum phase transitions. We start from and solve exactly a novel
Yang-Baxter spin-$1/2$ chain model with inhomogeneous and
anisotropic short-range interactions. For the ground state, we show
the behavior of neighboring entanglement in the parameter space and
find that the inhomogeneous coupling strengths affect entanglement
in a distinctive way from the homogeneous case, but this would not
affect the coincidence between entanglement and quantum criticality.
\end{abstract}
 \maketitle

\section{Introduction}\label{Sec-Intro}
Entanglement was conventionally considered to be a quirk of
microscopic objects and has been recognized to be ubiquitous and so
robust that it promises applications in the quantum communication
and computation like technologies \cite{2000Nielson}. In the last
few years, there was an increasing interest in entanglement in
macroscopic physical systems \cite{2008Amico,2008Vedral}.
Entanglement may lead to further insight into condensed matter
physics. For example, in statistical mechanics, given that quantum
phase transitions (QPTs) occur at absolute zero and are driven by
quantum fluctuations, entanglement may provide additional
correlations for QPTs \cite{2002Osborne,2002Osterloh} that have no
classical counterpart. In return, materials and experience built up
over the years in condensed matter are helping in finding new
protocols for quantum computation and communication.

Studying entanglement of the ground state in macroscopic physical
systems is crucial to understand a large variety of collective
quantum phenomena. For a one-dimensional spin-$1/2$ $XY$ chain
\cite{1958Anderson} with short-range interactions
\begin{equation}\label{eq-XY}
H_{XY}=-\sum_n\left(\frac{1+\gamma}{2}\sigma_n^x\sigma_{n+1}^x+\frac{1-\gamma}{2}\sigma_n^y\sigma_{n+1}^y+\lambda
\sigma_n^z\right),
\end{equation}
the entangled degree (ED) between any two nearest-neighbor particles
keeps the same for the translational symmetry, and its derivative is
capable to fulfill the role of an order parameter to characterize
QPT at the critical point $\lambda=1$
\cite{2003Vidal,2002Osborne,2002Osterloh}. Potential as it is, we
should ask whether such an observation is universal enough to assure
all correspondences between entanglement and QPT. For example, in
Ref. \cite{2006Campos} it demonstrates a long-distance entanglement
appearing for values of the microscopic parameters which do not
coincide with known quantum critical points. In addition, if the
short-range interactions are not homogeneous, e.g., a dual chain, ED
between two nearest-neighbor particles would generally not keep in
accordance for different sets of chains. At this case, questions
arise on what ED is and whether it well coincides with critical
points.

For self-contained and later use, let us recall the Yang-Baxter
approach for entangled states with two qubits \cite{2007Chen}. The
unitary $\breve{R}(\theta,\phi)$ matrix in this approach, taking
$\breve{R}(\theta) =\sin\theta+ \cos\theta\mathcal{M}(\phi)$ with
the generalized four-dimensional imaginary unit $\mathcal{M}$
($\mathcal{M}^2=-1$), is used to produce entangled states when
acting on direct-product states and EDs of the resulting states are
simply $|\sin2\theta|$. Supposing that $\theta$ is time-independent
and $\phi$ is time-dependent, we can get a Hamiltonian through
$H(\theta,\phi)=i\hbar(\partial \breve{R}/\partial t)
\breve{R}^\dag$ (see Eq. (\ref{eq-Hm})), which only governs the
evolution of entanglement varying with parameter $\theta$. At the
case of $\theta=\pi/2$, $\breve{R}$ corresponds to an identity
operation with no entanglement at all and $H$ thus vanishes. Such a
peculiar behavior of $H$ actually creates a nonanalytical point of
the ground state energy with respect to $\theta$ and this should be
reflected by some properties of its ground state such as the
geometric phase (GP). When extended to an infinite lattice, the
possibilities are richer and the vanishing point may correspond to a
QPT.

The purpose of this paper is twofold: one is that we solve exactly a
novel Yang-Baxter spin-$1/2$ chain model with alternating coupling
strengths by means of the Jordan-Wigner transformation and GP of the
ground state is examined for QPT; the other concerns the consequence
of entanglement between two local nearest-neighbor particles of the
chain, based on which we check whether entanglement under
inhomogeneous coupling strengths can well characterize critical
phenomena. This article is organized as follows. In Sec.
\ref{sec-model}, we introduce and exactly solve an inhomogeneous
Yang-Baxter spin-$1/2$ chain model. Based on the solution, we
investigate the quantum criticality by analyzing GP of the ground
state, and study the effect of inhomogeneity on entanglement between
different nearest-neighbor sites and next-nearest-neighbor sites in
Sec. \ref{sec-entangle}. At last, Sec. \ref{sec-concl} is dedicated
to the conclusion.

\section{The Yang-Baxter Spin-$1/2$ Chain Model and Exact Diagonalization\label{sec-model}}
The Yang-Baxter equation was originated from solving the
$\delta$-function interaction model by Yang \cite{1968Yang} and the
statistical models by Baxter \cite{1982Baxter}, and was then
introduced to solve many quantum integrable models by Faddeev and
Leningrad Scholars \cite{1980Sklyanin}. It plays a fundamental role
in the theories of 1+1 and 2+1 dimensional integrable quantum
systems, including lattice statistical models and nonlinear field
theory. For example, Yang's $\breve{R}$-matrix in YBE for the $n$th
and $(n+1)$th particles is $\breve{R}(u)_{n,n+1}=1+u\
\mathcal{P}_{n,n+1}$ ($u$ is a spectral parameter, i.e.,
one-dimensional momentum and $\mathcal{P}$ is permutation satisfying
$\mathcal{P}^2=1$); it yields the XXX chain model through
$H_{n,n+1}\sim
\partial\breve{R}/\partial u|_{u=0}$, when $\mathcal{P}$ takes its
four-dimensional representation of
$\mathcal{P}_{n,n+1}=\frac{1}{2}(1+{\bm \sigma}_n\cdot{\bm
\sigma}_{n+1})$ with ${\bm \sigma}$ being the Pauli matrix.
Considering the form of $\breve{R}(\theta_n)$ in Sec.
\ref{Sec-Intro}, it has a two-body interacting Hamiltonian
\cite{2007Chen}:
\begin{eqnarray}\label{eq-Hm}
H_{n,n+1}&=&-\hbar\omega\cos\theta_n\big[\cos\theta_n\big(S_n^z+S_{n+1}^z\big)\nonumber\\
&+&\sin\theta_n\big(e^{i\phi}S_n^+S_{n+1}^++e^{-i\phi}S_n^-S_{n+1}^-\big)\big],
\end{eqnarray}
where $S^\pm=S^x\pm iS^y$ with $S^{x,y,z}=\sigma^{x,y,z}/2$, and
$\phi$ is the flux dependent of time $t$ and it takes
$\phi(t)=\omega t$, denoting procession angle of spins around the
$z$ direction in a rotating magnetic field. For many particles, we
should sum all of them as $H=\sum_n H_{n,n+1}$ and if all of
$\theta_n$ are taken to be the same, it would correspond to a
homogeneous chain, otherwise it would correspond to an inhomogeneous
one.

From Eq. (\ref{eq-Hm}), the family of Hamiltonians that is
parameterized by $\phi$ is clearly isospectral, and, therefore, the
critical behavior is independent of $\phi$. In fact, we can see that
the spin raising-raising or lowering-lowering structure in Eq.
(\ref{eq-Hm}) allows a rotation for each spin around $z$-axis and
such a rotation transformation can be employed to adjust the value
of phase factors in Eq. (\ref{eq-Hm}), e.g.,
$\mathcal{H}=g(\phi/2)Hg^\dag(\phi/2)$ and
$g(\phi)=\prod_{l=1}^Ne^{-i\sigma_l^z\phi/2}$ giving
$g(\phi/2)S^+_ng^\dag(\phi/2)=e^{-i\phi/2}S_n^+$ and
$g(\phi/2)S^-_ng^\dag(\phi/2)=e^{i\phi/2}S_n^-$.  Then the
Hamiltonian is reduced to
\begin{eqnarray*}
\mathcal{H}&=&-\frac{1}{2}\hbar\omega\sum_{n}\cos\theta_n\big[\sin\theta_n(\sigma_n^x\sigma_{n+1}^x-\sigma_n^y\sigma_{n+1}^y)\\
&&+\cos\theta_n(\sigma_n^z+\sigma_{n+1}^z)\big].
\end{eqnarray*}
Comparing $\mathcal{H}$ with the $XY$ chain in Eq. (\ref{eq-XY}),
one sees that the structure of dominant two-body interaction in
$\mathcal{H}$ is exactly that in $H_{XY}$ yet under
$\gamma,\lambda\gg 1$ limit. Like $H_{XY}$, there is a global $Z_2$
symmetry for $H$ that keeps it invariant under a unitary
transformation $\prod_n\sigma_n^z$. In the following, we will see
that the ground state does not break such a symmetry.

By noting the fact that it can define these operators:
$J^z_n=\frac{1}{2}(S_n^z+S_{n+1}^z)$,
$J^x_n=\frac{1}{2}(S_n^+S_{n+1}^++S_n^-S_{n+1}^-)$ and
$J_n^y=\frac{i}{2}(S_n^-S_{n+1}^--S_n^+S_{n+1}^+)$ that satisfy the
angular momentum commutation relation, we see that ${\bm J}_n$
actually only occupies a subspace spanned by
$|\uparrow\uparrow\rangle_{n,n+1}$ and
$|\downarrow\downarrow\rangle_{n,n+1}$ that belongs to a $j=1/2$
angular momentum representation. Thus we can write the Hamiltonian
into an effective NMR-like form
$H_{n,n+1}=-\mathcal{B}_n(t)\cdot{\bm J}_n$, where the magnetic
field is
$\mathcal{B}_n(t)=2\hbar\omega\cos\theta_n(\sin\theta_n\cos\phi(t),-\sin\theta_n\sin\phi(t),\cos\theta_n)$.
Its eigenvalues are readily given by
$E^\pm_{n,n+1}=\pm\hbar\omega\cos\theta_n$ and eigenstates in
accordance are
$|E^\pm\rangle=\cos\frac{\theta_n}{2}|\uparrow\uparrow\rangle\pm\sin\frac{\theta_n}{2}e^{-i\phi}|\downarrow\downarrow\rangle$,
both EDs of which are $|\sin\theta_n|$. In the vicinity of
$\theta_n=\pi/2$, we can set $\cos\theta_n=\delta$ and see that it
keeps the forms of the eigenstates $|E^\pm(\delta)\rangle$ and hence
EDs even when $\delta\rightarrow 0$.

\begin{figure}
\includegraphics[width=8cm]{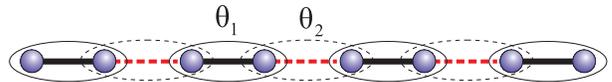}\\
\caption{A chain model with two-body interactions having the
Hamiltonian equation (\ref{eq-H}): the parameters $\theta_1$ on all
solid dark lines and $\theta_2$ on all dashed red lines describe two
different coupling strengths; each ellipse including solid and
dashed ones represents a two particle composite qubit with the
angular momentum ${\bm J}_i$. \label{fig1}}
\end{figure}

Now, let us extend Eq. (\ref{eq-Hm}) to an inhomogeneous chain which
approaches infinite when the particle number takes to be arbitrarily
large. For simplicity, we confine bonds between any pair of odd-even
numbered nearest-neighbor sites to be the same and characterized by
parameter $\theta_1$ and that between any pair of even-odd numbered
nearest-neighbor sites to be the same and characterized by
$\theta_2$ (see Fig. \ref{fig1}). By requiring the periodical
boundary condition, there are totally $2N$ sites and the Hamiltonian
can be written as
\begin{eqnarray}\label{eq-H}
H&=&-\sum_{n=1}^{N}(\mathcal{B}_1\cdot{\bm
J}_{2n-1}+\mathcal{B}_2\cdot{\bm J}_{2n}),
\end{eqnarray}
which can be interpreted simply as two sets of spins in an external
field with different coupling strengths along $z$-axis. The
composite qubit ${\bm J}_i$ satisfies $[{\bm J}_i,{\bm J}_j]=0$ for
$|i-j|\geq2$. At the same time, the property of $H$---zero
interaction at zero field---provides a critical phenomenon at
$\theta_1=\theta_2=\pi/2$ and is similar but not exactly the same to
the on-site exchange interactions \cite{2007Anderlini} and
superexchange interactions \cite{2008Trotzky} with ultracold atoms
in optical lattices, which would vanish as the optical field is
taken off. Here we will focus more on the entangled aspect of these
interactions by using the Hamiltonian Eq. (\ref{eq-H}), which has a
closed relation with entangled states. As for this, it is
interesting to ask what ED is when extending to an infinite chain,
especially for an inhomogeneous chain and whether it can effectively
determine the critical point.

To solve the model, let us first introduce the following
Jordan-Wigner transformation to represent spin operators at sites
with spinless fermion operators: $a_n=(\prod_{l<n}\sigma_l^z)S_n^+$
or $S_n^+=a_ne^{-i\pi\sum_{l<n}a^\dag_la_l}$. Then, note that the
summation in Eq. (\ref{eq-H}) is either on all even or on all odd
indices, and so we should distinguish even and odd to define their
corresponding forms in momentum space, respectively, as
\begin{equation}
a_k^e=\frac{1}{\sqrt{N}}\sum_{m=1}^Ne^{-i\frac{2\pi k}{N}m}a_{2m},\
a_k^o=\frac{1}{\sqrt{N}}\sum_{m=1}^Ne^{-i\frac{2\pi k}{N}m}a_{2m-1},
\end{equation}
where the reduced momentum $k=-M,\ldots,M$ with $M=(N-1)/2$ for $N$
odd and fermion operators $(a_k^{e(\dag)},a_k^{o(\dag)})$
anticommute with each other. Thus the Eq. (\ref{eq-H}) can be
written into
\begin{eqnarray}
H&=&-\frac{1}{2}\hbar\omega\sum_{k=-M}^M\big[(\xi_k e^{i\phi}
a_k^oa_{-k}^e+h.c.)\nonumber\\
&&-\Delta(a_k^{o\dag}a_k^o+a_k^{e\dag}a_k^e-1)\big],
\end{eqnarray}
with $\xi_k=\sin2\theta_2e^{2i\pi k/N} -\sin2\theta_1$ and
$\Delta=\cos2\theta_1+\cos2\theta_2+2$. The Hamiltonian $H$ can be
diagonalized by using the Bogoliubov transformation and the result
is
\begin{equation}\label{eq-Hdiag}
H=\frac{1}{2}\hbar\omega\sum_k\varepsilon_k^{\pm}(\alpha_k^\dag\alpha_k+\beta_k^\dag\beta_k-1).
\end{equation}
The eigenspectra contain two bands of quasiparticle excitations:
$\varepsilon_k^\pm=\pm\sqrt{|\xi_k|^2+\Delta^2}$. 
The transformed fermion operators $\alpha_k=u_k e^{i\phi/2}a_k^o+v_k
e^{-i\phi/2} a_{-k}^{e\dag}$ and $\beta_k=\bar{u}_ke^{i\phi/2}
a_k^e+\bar{v}_k e^{-i\phi/2}a_{-k}^{o\dag}$, where
$\bar{u}_k=-u_k=(\Delta+\varepsilon_k^\pm)/[2\varepsilon_k^\pm(\Delta+\varepsilon_k^\pm)]^{1/2}$
and
$\bar{v}_k=v_k^\ast=-\xi_k/[2\varepsilon_k^\pm(\Delta+\varepsilon_k^\pm)]^{1/2}$
for different bands $\varepsilon_k^\pm$. For these coefficients,
there is $\bar{u}_ku_k^\ast+\bar{v}_kv_k^\ast=0$.

The ground state $|g\rangle$ of $H$ is the vacuum of the fermionic
modes, satisfying $\alpha_k|g\rangle=0$ and $\beta_k|g\rangle=0$ for
all $k$. Generally it is hard to write the ground state obviously
into a spin superposed state, but for our model there is a fact that
the \emph{N\'{e}el state}---
$|\Psi^1\rangle=|\uparrow\downarrow\rangle^{\otimes N}$ or
$|\Psi^2\rangle=|\downarrow\uparrow\rangle^{\otimes N}$--- just
corresponds to the zero energy eigenstate of $H$. Take
$|\Psi^1\rangle$ for example: it has $\alpha_k|\Psi^1\rangle=0$ and
therefore it is identical to
$|\Psi^1\rangle=\prod_k\beta_k^\dag|g\rangle$, which inversely gives
an expression for the ground state
$|g\rangle=\prod_k\beta_k|\Psi^1\rangle$, i.e.,
\begin{eqnarray}
|g\rangle&=&\prod_k\Big\{\sum_{m=1}^N\frac{e^{-i\frac{2\pi
k}{N}m}}{\sqrt{N}}\Big[\bar{v}_ke^{-i\phi/2}\Big(\prod_{l<2m-1}\sigma_l^z\Big)S_{2m-1}^-\nonumber\\
&&+\bar{u}_ke^{i\phi/2}\Big(
\prod_{l<2m}\sigma_l^z\Big)S_{2m}^+\Big]\Big\}|\uparrow\downarrow\rangle^{\otimes
N},\label{eq-g1}
\end{eqnarray}
which would return to the biparticle case (i.e., $|E^-\rangle$) if
one takes $N=1$ and $\theta_2=\pi/2$. When $\theta_1\neq \pi/2$ and
$\theta_2=\pi/2$, we can see that Eq. (\ref{eq-H}) becomes to
describe $N$ isolated dimers, which has an exact ground state,
$\prod_{m}(\cos\frac{\theta_1}{2}|\uparrow\uparrow\rangle_{2m-1,2m}+\sin\frac{\theta_1}{2}e^{-i\phi}|\downarrow\downarrow\rangle_{2m-1,2m})$.
From Eq. (\ref{eq-g1}), it can be seen the ground state is invariant
under the global $Z_2$ transformation and so it keeps the same
symmetry as the Hamiltonian. Alternatively, the ground state can
also be expressed by
\begin{equation}\label{eq-g}
|g\rangle=\prod_k\left(\bar{u}_ke^{i\phi/2}|0\rangle_{-k}^o|0\rangle_k^e+\bar{v}_ke^{-i\phi/2}|1\rangle_{-k}^o|1\rangle_k^e\right),
\end{equation}
where $|0\rangle^{o,e}_k$ and $|1\rangle^{o,e}_k$ are the vacuum and
single excitation of the $k$th mode, $a_k^{o,e}$, respectively. The
ground state is a tensor product of states, each lying in the
two-dimensional Hilbert space spanned by
$|0\rangle_{-k}^o|0\rangle_{k}^e$ and
$|1\rangle_{-k}^o|1\rangle_{k}^e$. Such a form provides us a
convenient way to discuss its dynamical property such as GP of the
ground state.

\section{Quantum Criticality and entanglement}\label{sec-entangle}
QPT occurs at a point in the external parameter space, where there
can be a level-crossing and excited levels become the ground state,
creating a point of nonanalyticity of the ground state energy as a
function of external parameters \cite{1999Sachdev}. With the
Hamiltonian Eq. (\ref{eq-Hdiag}) in consideration, we take
$\theta_{1,2}$ as those external parameters. Obviously, at the point
of $\theta_1=\theta_2=\pi/2$ all energy levels cross and hence it is
a critical point, but it is different from the conventional QPT by
having a vanishing Hamiltonian and for convenience of discussion, we
might as well call it a QPT. Recently, GP of the ground state
\cite{2005Carollo,2006Zhu} and ED between nearest-neighbor particles
\cite{2002Osborne,2002Osterloh} for a homogeneous Heisenberg XY
chain were proposed to characterize the criticality of QPT. In this
section, we investigate GP and ED for our novel inhomogeneous chain,
 analyze their behaviors as the parameters $\theta_{1,2}$ vary,
 and further discuss their nonanalytical property in the proximity of
 QPT.

\subsection{Quantum Criticality Characterized by
Geometric Phase} GP of the ground state, accumulated by varying the
angle $\phi$ from $0$ to $2\pi$, is described by
$\beta_g=\frac{1}{N}\int_0^{2\pi}\langle g|i\partial_{\phi}|g\rangle
d\phi$, and by utilizing Eq. (\ref{eq-g}) it is
\begin{equation}\label{eq-GP}
\beta_g^\pm=-\frac{\pi}{N}\sum_k(|\bar{u}_k|^2-|\bar{v}_k|^2)=-\frac{\pi}{N}\sum_k\Delta/\varepsilon_k^\pm,
\end{equation}
with $\beta_g^+=-\beta_g^-$. The term
$\beta_k=-\pi\Delta/\varepsilon_k^\pm$ is a geometric phase for the
$k$th mode, and represents the area in the parameter space enclosed
by the loop determined by $(\theta_1,\theta_2,\phi)$. One can see
that when we turn off the coupling between dimers by setting
$\theta_2=\pi/2$, GP would return to the biparticle dimer case with
$\beta_g=\pi(1-\cos\theta_1)$. To study quantum criticality, we are
interested in the thermodynamic limit when the spin lattice number
$N\rightarrow \infty$. In this case the summation
$\frac{1}{N}\sum_{k=-M}^{M}$ can be replaced by the integral
$\frac{1}{\pi}\int_0^{\pi}d\varphi$ with $\varphi=2\pi k/N$; GP in
the thermodynamic limit is given by
\begin{equation}
\beta_g^\pm=-\int_0^\pi d\varphi \Delta/\varepsilon_\varphi^\pm,
\end{equation}
where the energy spectra
$\varepsilon_\varphi^\pm=\pm\sqrt{|\xi_\varphi|^2+\Delta^2}$ with
$|\xi_\varphi|^2=\sin^22\theta_1+\sin^22\theta_2-2\sin2\theta_1\sin2\theta_2\cos\varphi$.

\begin{figure}
\includegraphics[width=8cm]{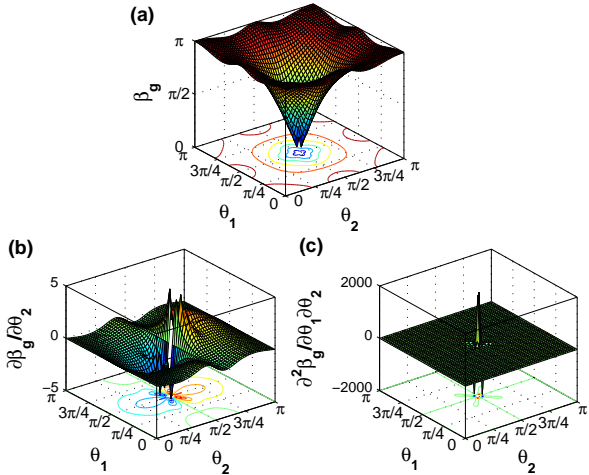} \\
\caption{(Color online) The diagram for the Berry phase of the
ground state of the spin-$1/2$ chain model: (a) the Berry phase
$\beta_g^+$ corresponding to the ground state; (b) its derivative
$\partial\beta_g^+/\partial\theta_2$ as a function of $\theta_1$ and
$\theta_2$; (c) its derivative
$\partial^2\beta_g^+/\partial\theta_1\partial\theta_2$ as a function
of $\theta_1$ and $\theta_2$. }\label{fig2}
\end{figure}

To see the quantum criticality obviously, we plot GP $\beta_g$ and
its derivatives $\partial\beta/\partial{\theta_2}$,
$\partial^2\beta/\partial{\theta_1}\partial{\theta_2}$ in the
parameter $(\theta_1,\theta_2)$ space, shown in Fig. \ref{fig2}. It
can be seen, from Fig. \ref{fig2}(a), that there is a conical
intersection at the point $\theta_1=\theta_2=\pi/2$, which indicates
a nonanalytical point there. The nonanalytical property at the
critical point can be seen obviously from the diagram of GP
derivative [see Fig. \ref{fig2}(c)]. However, as we pointed above,
such a QPT point is trivial, since at the point the whole
Hamiltonian vanishes and hence it appears to be exotic there. If
fixing one parameter, say $\theta_1$, it would correspond to the
uniparameter case and we should check whether there are other
critical phenomena as varying $\theta_2$. The derivative of GP with
$\theta_2$ is plotted in Fig. \ref{fig2}(b), from which we can see,
except for the above critical point and its vicinity, it is analytic
everywhere. So there is no additional critical point.

\subsection{Entanglement of the Ground State} In
this section, we confine our interest at entanglement between
nearest-neighbor sites in the chain, given long-distance
entanglement decays rapidly with the distance (see below). To
describe entanglement, we use the concurrence \cite{1998Wooters} of
a biparticle state, related to the ``entanglement of formation"
\cite{1996Bennett}, to define ED of a state. The concurrence for the
state of the $i$th and $j$th particles is defined as
\begin{equation}\label{eq-concurrence}
C(i,j)=\max\{r_1-r_2-r_3-r_4, 0\},
\end{equation}
where $r_{1,2,3,4}$ are the square roots of the eigenvalues of the
product matrix $R=\rho(i,j)\tilde{\rho}(i,j)$ in descending order;
$\rho(i,j)$ is the density matrix of the $i$th and $j$th spin-$1/2$
particles and the spin flipped matrix is defined as
$\tilde{\rho}(i,j)=\sigma^y\otimes\sigma^y\rho^\ast(i,j)\sigma^y\otimes\sigma^y$.
If it is a pure state, e.g. $|E^-\rangle$, the density matrix
$\rho(n,n+1)=|E^-\rangle\langle E^-|$ and the concurrence
quantifying entanglement is $C(n,n+1)=|\sin\theta_n|$. If it is a
biparticle state in a multiparticle system, $\rho(i,j)$ would
represent a biparticle mixed state reduced from the multiparticle
density matrix $\rho$.

For the chain in consideration, translation invariance of dual
lattices implies that $C(2m, 2m+1)=C_e(1)$, $C(2m-1,2m)=C_o(1)$ and
$C(2m-1,2m+1)=C(2)$ for all $m$. The concurrence will be evaluated
as a function of the relative position $|i-j|$ between the $i$th and
$j$th spins and parameters $\theta_{1,2}$. All information needed is
contained in the reduced density matrix $\rho(i,j)$ obtained from
the ground-state wavefunction after all the spins except those at
positions $i$ and $j$ have been traced out. The resulting
$\rho(i,j)$ represents a mixed state of a biparticle system. The
structure of the reduced density matrix is obtained by exploiting
symmetries of the chain. The nonzero entries of $\rho(i,j)$ can then
be related to the various correlation functions
\cite{1971Barouch,1961Lieb,1970Pfeuty} as
\begin{eqnarray}
\rho(i,j)&=&\Big(I+\langle\sigma_i^z\rangle \sigma_i^z\otimes
1+\langle\sigma_j^z\rangle
1\otimes\sigma_j^z+\langle\sigma_i^z\sigma_j^z\rangle
\sigma_i^z\otimes\sigma_j^z\nonumber\\&&+\sum_{X,Y=x,y}\langle\sigma_i^X\sigma_j^Y\rangle
\sigma_i^X\otimes\sigma_j^Y\Big)/4.
\end{eqnarray}

Correlation functions under the ground state can be evaluated by
using the fermionic representation and the simple identity
$1-2a_n^\dag a_n=(a_n^\dag+a_n)(a_n^\dag-a_n)$. For each pair of
fermion operators $(a,a^\dag)$, we can further define two Majorana
fermion operators $(A,B)$: $A_n=a_n^\dag+a_n$ and
$B_n=i(a_n^\dag-a_n)$ with $A^\dag=A$ and $B^\dag=B$. Exploring
them, we can write the Pauli matrices as:
$\sigma_n^x=\prod_{l<n}[(-i)A_lB_l]A_n$,
$\sigma_n^y=\prod_{l<n}[(-i)A_lB_l]B_n$, and
$\sigma_n^z=(-i)A_nB_n$. A two-body correlation function, say
$\langle \sigma_m^x\sigma_{n}^x\rangle|_{m<n}$, under the ground
state, is
\begin{eqnarray}
\lefteqn{\langle \sigma_m^x\sigma_n^x\rangle=-i\left\langle
B_m\prod_{l=m+1}^{n-1}(-iA_lB_l)A_n\right\rangle}\\
&=&\langle (-iB_m)A_{m+1}(-iB_{m+1})\cdots
A_{n-1}(-iB_{n-1})A_{n}\rangle.\nonumber
\end{eqnarray}
Since the expectation values are with respect to a free Fermi
theory, the expression on the right-hand side can be evaluated by
the Wick's theorem \cite{1999Sachdev,2004Wen}, which relates it to a
sum over products of expectation values of pairs of operators, i.e.,
$\langle A_lA_m\rangle$, $\langle B_lB_m\rangle$, and $\langle
B_lA_m\rangle$. The evaluation of average values of these pairs is
displayed in Appendix \ref{sec-Cal}. In order to see the macroscopic
property of entanglement, we define two $k$-independent functions:
\begin{eqnarray}\label{eq-FG}
\mathcal{F}(|n-m|)&=&\frac{1}{N}\sum_ke^{i\frac{2\pi
k}{N}(n-m)}(|\bar{u}_k|^2-|\bar{v}_k|^2),\nonumber\\
\mathcal{G}(m-n)&=&\frac{1}{N}\sum_ke^{-i\frac{2\pi
k}{N}(n-m)}2u_kv_k,
\end{eqnarray}
which have summed all frequencies in the momentum space to be the
form in the position representation. We can see, $\mathcal{F}(0)$ is
nothing but the one proportional to GP in Eq. (\ref{eq-GP}). In this
respect, we may well say $\mathcal{F},\mathcal{G}$ are macroscopical
quantities, which under the thermodynamical limit
$N\rightarrow\infty$ can be calculated still by making the
replacement $\frac{1}{N}\sum_k\rightarrow \frac{1}{\pi}\int_0^\pi
d\varphi$.

\begin{figure}
\includegraphics[width=8cm]{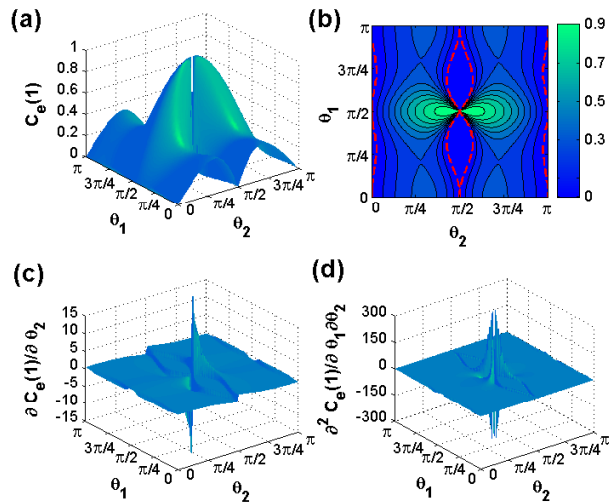}\\
\caption{(Color online) The diagram for concurrence in the parameter
space: (a) ED in the form of concurrence is shown and $C_e(1)$
denotes entanglement between even-odd numbered two nearest-neighbor
sites; Its corresponding contours are also shown in (b); (c) and (d)
display the first derivative $\partial C_e(1)/\partial \theta_2$ and
the second derivative
$\partial^2C_e(1)/\partial\theta_1\partial\theta_2$ of $C_e(1)$,
respectively. }\label{fig3}
\end{figure}

Next, we would focus on entanglements between the nearest-neighbor
sites. In principle, the numerical results on the concurrence of
such two-site density state can be performed readily according to
its definition introduced above, but before that, we find the
concurrence depends only on the above two $k$-independent functions
$\mathcal{F}(|n-m|)$ and $\mathcal{G}(m-n)$. As an illustration, we
give out the expressions of ED in the form of concurrence between
odd-even and even-odd neighboring sites, respectively, by
\begin{eqnarray}
C_o(1)&=&\max\{0,|\mathcal{G}(0)|-\frac{1}{2}|\mathcal{F}(0)^2+\mathcal{G}(0)^2-1|\},\nonumber\\
C_e(1)&=&\max\{0,|\mathcal{G}(1)|-\frac{1}{2}|\mathcal{F}(0)^2+\mathcal{G}(1)^2-1|\}.
\end{eqnarray}
Their varying trends in the parameter space are displayed in Fig.
\ref{fig3}(a) associated with contours in \ref{fig3}(b). Note that
although we plot merely the concurrence of even-odd neighboring
sites in Fig. \ref{fig3}, that of odd-even neighboring sites can be
immediately gotten by noticing the symmetry:
$C_o(\theta_1\leftrightarrow\theta_2)=C_e$.

Fig. \ref{fig3}(a) shows that the maximum of $C_e(1)$ (or $C_o(1)$)
approaches but cannot reach one (i.e, the intrinsic maximal value of
ED) because of the vanishing Hamiltonian at the value. At
$\theta_1=\pi/2$ even-odd pairs become decoupled from the rest and
the behavior of $C_e$ for $\theta_2\rightarrow\pi/2$ is the same as
in the biparticle case. Whereas, the difference refers to the
minimum of $C_e(1)$ that relates to disentangled states with
vanishing ED and from Fig. \ref{fig3}(b) we can see that in the
parameter space it occupies a considerable regime enclosed by the
dashed red lines for disentangled states. Thereby, at the interface
of the entangled and disentangled states it will correspond to the
unsmoothed part of ED mainly due to the unsmoothed definition of the
concurrence [see Eq.(\ref{eq-concurrence})] and the inhomogeneous
interaction strength in the chain, which would as well imply the
nonanalytical behavior for derivatives of ED with respect to
$\theta_{1,2}$. In contrast, at the homogeneous case of $C_e=C_o$
achieved along the diagonal line of either $\theta_1=\theta_2$ or
$\theta_1+\theta_2=\pi$ in Fig. \ref{fig3}(b), ED has only one
unsoomthed point which corresponds to the critical point
appropriately.

Fig. \ref{fig3}(c) and (d) show respectively the first and second
derivatives of ED with respect to parameters $\theta_{1,2}$. The
dominating divergence at the central region is well in agreement to
that appeared in the derivatives of GP and thus indicates QPT.
Except this, it does not diverge at other points, exhibiting instead
just a finite discontinuity at the border between separable and
pairwise entangled sectors. This type of non-analytic behavior is
quite distinct from that at the critical point and stems just from
the definition of concurrence. In addition, the divergence includes
both upward and downward directions and this reflects the increasing
and decreasing trends of ED in the vicinity of the nonanalytical
region.

As for two next-nearest-neighbor sites and other farther neighboring
sites, ED  should decay rapidly with the distance (generally even
more rapidly than standard correlations) for the short-ranged
interaction. To illustrate it, our numerical result demonstrates
that the next-nearest-neighbor concurrence $C(2)$ vanishes for the
whole chain. As a result, it is sufficient for us to only take the
nearest-neighbor biparticle entanglement into account on this chain,
while for multiparticle entanglement it may has a connection with
the so called topological quantum phase transition (e.g.,
\cite{2005Hamma}) but not discussed here.

\section{conclusion and discussion}\label{sec-concl}
To summarize, we have demonstrated an exact solution to a particular
spin-$1/2$ chain model with alternating nearest-neighbor coupling
strengths and have analyzed the influence of inhomogeneous
interaction on the ground state through GP and ED approaches. By
evaluating GP of the ground state, we display its behavior at the
parameter space, from which a critical point could be determined
through the divergent derivative of GP. After that, via examining
the biparticle entanglement by virtue of concurrence, we also show
the tendency of ED with respect to parameters and find that ED and
its derivatives can determine the critical point as well as GP does
at the inhomogeneous case. Although ED has an unsmoothed definition
at the border of separable and pairwise entangled sectors, it does
not exactly affect the nearest-neighbor concurrence to detect the
critical point, exhibiting instead a clear signature of it in its
derivative as seen in Fig. \ref{fig3}(c).

As remarked earlier, the specific parametrization of the Hamiltonian
has an intimate relation with ED at the biparticle case and when
extending it to an infinite lattice, the result is interesting: for
homogeneous coupling strengths, ED has suppressed values with its
maximum far less than one, keeping equal for every pair of
nearest-neighbor sites; for inhomogeneous coupling strengths, ED
appears to have different values between even-odd numbered and
odd-even numbered nearest-neighbor sites, from which a very high ED
is available (see Fig. \ref{fig3}). In a way, the property might
apply to other inhomogeneous lattice models as a manifestation of
general principles. Also, the analysis of the inhomogeneous
entanglement for a condensed matter system is possibly of great
importance for creating ideal entanglement resources in quantum
information processing.

\begin{acknowledgments}
We thank J. L. Chen for helpful discussions. This work was supported
by NSF of China (Grants No. 10575053 and No. 10605013) and LuiHui
Center for Applied Mathematics through the joint project of Nankai
and Tianjin Universities.
\end{acknowledgments}

\appendix
\section{Calculation of expectation values of pairs of operators\label{sec-Cal}}
In determining the expectation values of pairs of Majorana operators
under the ground state, we have used the representation
$(\alpha_k,\beta_k)$ and the definition of the ground state
$\alpha_k|g\rangle=\beta_k|g\rangle=0$. If we define two independent
real functions, i.e., Eq. (\ref{eq-FG}) by
\begin{eqnarray*}
\mathcal{F}(|n-m|)&=&\frac{1}{N}\sum_ke^{i\frac{2\pi
k}{N}(n-m)}(|\bar{u}_k|^2-|\bar{v}_k|^2)\nonumber\\
&=&\frac{1}{N}\sum_k\cos[\frac{2\pi}{N}k(n-m)]\Delta/\epsilon_k^\pm,
\end{eqnarray*}
\begin{eqnarray*}
\mathcal{G}(m-n)&=&\frac{1}{N}\sum_ke^{-i\frac{2\pi
k}{N}(n-m)}2u_kv_k\nonumber\\
&=&\frac{1}{N}\sum_k\{\sin2\theta_2\cos[\frac{2\pi}{N}k(m-n-1)]
-\nonumber\\&&\sin2\theta_1\cos[\frac{2\pi}{N}k(m-n)]\}/\epsilon_k^\pm,\nonumber
\end{eqnarray*}
which are obtained under the thermodynamical limit
$N\rightarrow\infty$ by making the replacement
$\frac{1}{N}\sum_k\rightarrow \frac{1}{\pi}\int_0^\pi d\varphi$,
then the average values of Majorana operator pairs can be written as
\begin{eqnarray*}
&\langle A_{2n}A_{2m}\rangle=\langle
A_{2n-1}A_{2m-1}\rangle=\delta_{n,m},&\nonumber\\
&\langle B_{2n}B_{2m}\rangle=\langle
B_{2n-1}B_{2m-1}\rangle=\delta_{n,m},&\\
&\langle A_{2n}(-iB_{2m})\rangle=\langle
A_{2n-1}(-iB_{2m-1})\rangle=\mathcal{F}(|n-m|), \\
&\langle A_{2n}A_{2m-1}\rangle=\langle
(-iB_{2n})(-iB_{2m-1})\rangle=i\sin\phi \mathcal{G}(m-n),\\
&\langle A_{2n}(-iB_{2m-1})\rangle=\langle(-i
B_{2n})A_{2m-1}\rangle=\cos\phi \mathcal{G}(m-n).
\end{eqnarray*}
From these average values, we can calculate all correlation
functions by using Wick theorem. For the nearest-neighbor case of
$\rho(2m-1,2m)$, we have
\begin{eqnarray*}
\langle\sigma_{2m-1}^x\sigma_{2m}^x\rangle&=&\langle(-iB_{2m-1})A_{2m}\rangle=-\cos\phi
\mathcal{G}(0),\\
\langle\sigma_{2m-1}^y\sigma_{2m}^y\rangle&=&-\langle
A_{2m-1}(-iB_{2m})\rangle=\cos\phi
\mathcal{G}(0),\\
\langle\sigma_{2m-1}^x\sigma_{2m}^y\rangle&=&\langle\sigma_{2m-1}^y\sigma_{2m}^x\rangle=\sin\phi
\mathcal{G}(0),\\
\langle \sigma_{2m-1}^z\sigma_{2m}^z\rangle&=&\langle
A_{2m-1}(-iB_{2m-1})A_{2m}(-iB_{2m})\rangle\\
&=&\mathcal{F}(0)^2+\mathcal{G}(0)^2.
\end{eqnarray*}
For the nearest-neighbor case of $\rho(2m,2m+1)$, we have
\begin{eqnarray*}
\langle\sigma_{2m}^x\sigma_{2m+1}^x\rangle&=&\langle(-iB_{2m})A_{2m+1}\rangle=\cos\phi
\mathcal{G}(1),\\
\langle\sigma_{2m}^y\sigma_{2m+1}^y\rangle&=&-\langle
A_{2m}(-iB_{2m+1})\rangle=-\cos\phi
\mathcal{G}(1),\\
\langle\sigma_{2m}^x\sigma_{2m+1}^y\rangle&=&\langle\sigma_{2m}^y\sigma_{2m+1}^x\rangle=-\sin\phi
\mathcal{G}(1),\\
\langle \sigma_{2m}^z\sigma_{2m+1}^z\rangle&=&\langle
A_{2m}(-iB_{2m})A_{2m+1}(-iB_{2m+1})\rangle\\
&=&\mathcal{F}(0)^2+\mathcal{G}(1)^2.
\end{eqnarray*}
At last, for the next-nearest-neighbor case of $\rho(2m-1,2m+2)$, we
have
\begin{eqnarray*}
\langle\sigma_{2m-1}^x\sigma_{2m+1}^x\rangle&=&\langle(-iB_{2m-1})A_{2m}(-iB_{2m})A_{2m+1}\rangle\nonumber\\
&=&-\mathcal{G}(0)\mathcal{G}(1)-\mathcal{F}(0)\mathcal{F}(1),\\
\langle\sigma_{2m-1}^y\sigma_{2m+1}^y\rangle&=&-\langle
A_{2m-1}A_{2m}(-iB_{2m})(-iB_{2m+1})\rangle\nonumber\\
&=&-\mathcal{G}(0)\mathcal{G}(1)-\mathcal{F}(0)\mathcal{F}(1),\\
\langle\sigma_{2m-1}^x\sigma_{2m+1}^y\rangle&=&\langle\sigma_{2m-1}^y\sigma_{2m+1}^x\rangle=0,\\
\langle \sigma_{2m-1}^z\sigma_{2m+1}^z\rangle&=&\langle
A_{2m-1}(-iB_{2m-1})A_{2m+1}(-iB_{2m+1})\rangle\\
&=&\mathcal{F}(0)^2-\mathcal{F}(1)^2.
\end{eqnarray*}

\end{document}